\documentclass[12pt,preprint]{aastex}
\usepackage{graphicx}
\usepackage{epstopdf}

\usepackage{natbib}

\begin{document}

\title{Transit Timing Variation Analysis of OGLE-TR-132b with Seven New Transits\altaffilmark{1}}

\author{E. R. Adams\altaffilmark{2,3}, M. L\'opez-Morales\altaffilmark{4,5}, J. L. Elliot\altaffilmark{3,6,7},  S. Seager\altaffilmark{3,6},  D. J. Osip\altaffilmark{8}}

\altaffiltext{1}{This paper includes data gathered with the 6.5 meter Magellan Telescopes located at Las Campanas Observatory, Chile.}
\altaffiltext{2}{Harvard-Smithsonian Center for Astrophysics, 60 Garden St., Cambridge, MA, 02138}
\altaffiltext{3}{Department of Earth, Atmospheric, and Planetary Sciences, Massachusetts Institute of Technology, 77 Massachusetts Ave., Cambridge, MA, 02139}
\altaffiltext{4}{Institut de Ci\`encies de l'Espai (CSIC-IEEC), Campus UAB, Facultat de Ci\`encies, Torre C5, parell, 2a pl, E-08193 Bellaterra, Barcelona, Spain}
\altaffiltext{5}{Visiting Investigator; Carnegie Institution of Washington, Department of Terrestrial Magnetism, 5241 Broad Branch Road NW, Washington, DC 20015-1305}
\altaffiltext{6}{Department of Physics, Massachusetts Institute of Technology, 77 Massachusetts Ave., Cambridge, MA, 02139}
\altaffiltext{7}{Lowell Observatory, 1400 W. Mars Hill Rd., Flagstaff, AZ 86001}
\altaffiltext{8}{Las Campanas Observatory, Carnegie Observatories, Casilla 601, La Serena, Chile}

\begin{abstract}

We report the results of the first transit timing variation (TTV) analysis of the very hot Jupiter OGLE-TR-132b, using ten transits collected over a seven-year period. Our analysis combines three previously published transit light curves with seven new transits, which were observed between February 2008 and May 2009 with the new MagIC-e2V instrument on the Magellan Telescopes in Chile. We provide a revised planetary radius of $R_p = 1.23 \pm 0.07$ $R_J$, which is slightly larger than, but consistent within the errors, of the previously published results. Analysis of the planet-to-star radius ratio, orbital separation, inclination and transit duration reveals no apparent variation in any of those parameters during the time span observed. We also find no sign of transit timing variations larger than $-108 \pm 49$ s, with most residuals very close to zero. This allows us to place an upper limit of 5-10 $M_{\oplus}$ for a coplanar, low-eccentricity perturber in either the 2:1 or 3:2 mean-motion resonance with OGLE-TR-132b. We similarly find that the data are entirely consistent with a constant orbital period and there is no evidence for orbital decay within the limits of precision of our data. 

\end{abstract}

\keywords{stars: planetary systems -- OGLE-TR-132}

\section{Introduction}

Precise timing of exoplanet transits over many years is a powerful technique to learn more about a planetary system. Transit timing variations, or TTVs, can arise from many potential scenarios, such as orbital precession, orbital decay, and perturbations by additional planets, small stellar companions, large moons or rings; these effects have been modeled in numerous papers over the last decade \citep{MiraldaEscude2002, Sasselov2003, Holman2005, Agol2005, Heyl2007, Ford2007, Simon2007, Kipping2009a, Kipping2009b, Levrard2009}.  

Recently, \citet{Holman2010} announced the first incontrovertible evidence of transit timing variations in the Kepler-9 system, where a pair of Saturn-mass planets are trapped in a 2:1 mean-motion resonance, producing strong, complementary deviations of over an hour in the midtimes of both transits. Other observational evidence is less clear cut, with recent claims of timing variations in WASP-3b \citep{Maciejewski2010a} and WASP-10b \citep{Maciejewski2010b} announced but unconfirmed, while a claimed detection of TTVs for OGLE-TR-111b has not held up \citep{Adams2010a, Diaz2008}. Additionally, there have been claims of a tentative detection of variations in the transit duration of GJ436b, increasing by roughly 3 minutes per year \citep{Coughlin2008}, consistent with the $<8~M_{\oplus}$ limit for a second planet in this system placed by transit timing \citep{Bean2008}. Finally, previous work by our group has found very tentative evidence that the orbital period of OGLE-TR-113b is decreasing at a rate of $-60\pm15$ ms yr$^{-1}$, though additional observations are needed to confirm this result \citep{Adams2010b}.

Though perhaps less exciting, the $non$-detection of transit timing variations is both important and expected. It has been suggested that systems with hot Jupiters are statistically less likely to be part of multi-planet systems \citep{Wright2009}, and transit timing limits can place tighter mass limits on potential nearby companions than other detection methods to date, with upper limits of a few Earth masses possible for companions in strong mean-motion resonances. With observations spanning many years, upper limits can also be placed on the rate of orbital decay of the known hot Jupiter, providing valuable direct observational constraints on theoretical models of tidal dissipation in stars. Currently tidal dissipation in stars is only estimated indirectly, through arguments about circularization timescales of binary stars in clusters \citep[see e.g.][]{Meibom2005, Ogilvie2007}. 

In this work we present the case of the hot Jupiter OGLE-TR-132b, with seven new light curves observed and no timing variations larger than  $-108 \pm 49$ s detected during observations spanning seven years.  In Section~\ref{section:obs} we describe the observations, data analysis and light curve fitting. Section~\ref{section:fitting} describes the modeling and improvements to the system parameters, and places limits on parameter variability between transits. Section~\ref{section:timing} discusses the timing constraints placed on the planetary system. We conclude in Section~\ref{section:conclusions}.

\section{Observations and photometry}
\label{section:obs}

OGLE-TR-132b was first announced as a planet candidate by \citet{Udalski2003} and confirmed to be of planetary mass by \citet{Bouchy2004}. The star, OGLE-TR-132, is one of the faintest transit-hosting stars (I=15.7; RA(J2000)=10:50:34.72, Dec(J2000)=--61:57:25.9). It is somewhat larger and hotter than the sun, with $M_* = 1.26 \pm 0.03~M_{\odot}$, $R_* = 1.34 \pm 0.08~R_{\odot}$, $T_{eff} = 6210 \pm 59 K$, and relatively high metallicity, $[Fe/H] = +0.37 \pm 0.07$ dex \citep{Bouchy2004}. No reliable estimate of the age of OGLE-TR-132 exists \citep{Melo2006}. Two high-quality follow-up light curves in $R$-band were obtained by \citet{Moutou2004} and \cite{Gillon2007}, and remain until this work the only published photometry of this planet aside from the discovery data. Combining those transits with the \citet{Bouchy2004} radial velocity observations, \citet{Gillon2007} derived a radius for this planet of $R_p = 1.18 \pm 0.07~R_{J}$, which is slightly larger, but consistent within the errors, than the value of $R_p = 1.13 \pm 0.08~R_{J}$ derived by Moutou et el. (2004). More recently, \citet{Southworth2008} reanalyzed the available data and derived a new radius for the planet,  $R_p = 1.25 \pm 0.16~R_{J}$, and identified this system as one that needs additional photometry to improve the light-curve parameters.

Since 2006 we have conducted a campaign to search for transit timing variations \citep{Adams2010PhD}, using the same instrument and observing strategy to minimize systematic differences between data sets. Results for OGLE-TR-111b and OGLE-TR-113b have been previously published \citep{Adams2010a, Adams2010b}. OGLE-TR-132 is our faintest target to date, but because of the short period of the planet ($P=1.69$ days) there are many transit opportunities. We observed seven full transits of OGLE-TR-132b between April 2008 and May 2009, using the dual-CCD instrument MagIC on the Magellan-Baade 6.5m telescope. All transits were observed in the Sloan $i'$ filter using MagIC-e2v; an eighth, partial transit was attempted with MagIC-SiTe in February 2008, but a faulty shutter rendered those data unusable. In the remainder of the paper we refer to each transit by the UT date when it was observed, following the notation YYYYMMDD.

The MagIC-e2v CCD saw first light on Magellan in April 2008. The camera has low readout noise (about 6 e- per pixel), a small field-of-view ($38\arcsec \times 38\arcsec$) and high-resolution pixels ($0.\arcsec037$ per pixel), which both minimizes blends and produces images of the stars spread over many pixels (FWHM=6--20 pixels, depending on seeing and binning). This latter feature reduces differential pixel response effects and increases the total number of photons that can be collected per frame, without requiring defocusing. The e2v gain was 2.4 e-/ADU in 2008 and 0.5 e-/ADU in 2009 (after engineering changes). The main advantage of the e2v over conventional CCDs is its frame-transfer capability: the readout time per frame is only 5s in standard readout mode and a few milliseconds in frame transfer mode, surpassing the 23s readout time of the MagIC-SITe chip.

Given that the main focus of our observations is accurate timing, we took special care to correctly record the time in the image headers. The 2008 MagIC-e2v times came from the network server, which we verified to be synchronized with the observatory's GPS clocks at the beginning of each night. The 2009 MagIC-e2v times came from an embedded PC104 computer, which receives unlabeled GPS pulses every second, and is synchronized to the observatory's GPS receiver every night. The intrinsic error for all header times is thus much less than a second.

The observational conditions for each transit are summarized in Table~\ref{table:ogle132obsparams}. Because MagIC-e2v is a new camera, we have observed with several different configurations in order to test performance. All three transits in 2008 were observed in the same configuration: 5s readout-mode and 1$\times$1 binning. Three transits in 2009 used 2$\times$2 binning, while one transit (20090424) was binned 1$\times$1; all 2009 transits used the frame-transfer mode with millisecond readout times. Exposure times ranged from 10-120 s depending on the binning and the weather conditions, and were adjusted with a goal of having at least $10^6$ counts on the target on each frame. A few images were discarded because of low counts or telescope tracking problems during the observations, as noted in Table~\ref{table:ogle132obsparams}. For calibration data, in 2008 we used twilight flats (5 per night), while in 2009 we switched to dome flats because we could get many more of them (typically 100 per night), thus reducing the potential noise introduced during the calibration stage. Although bias frames were taken each night, we did not use them, preferring instead to subtract the bias level recorded in the overscan region on each data frame.

The data for each night were overscan-corrected and flatfielded using standard IRAF routines\footnote{IRAF is distributed by the National Optical Astronomy Observatory, operated by the Association of Universities for Research in Astronomy, Inc., under cooperative agreement with the National Science Foundation}. Simple aperture photometry, using the IRAF {\it apphot} package, provided optimal results, since both the target and several comparison stars appear well isolated in all the frames.  We ran a wide grid of apertures and sky annuli to find the values which provided the optimal photometry, i.e. the values which minimized the out-of-transit photometric dispersion of each light curve. The comparison stars were iteratively selected to be 1) similar in brightness to the target, 2) photometrically stable, and 3) un-blended. Of the 10--20 stars examined for each transit, we selected between 2--8 as final photometric comparisons, depending on the night. The best aperture radii ranged between 9--26 pixels, while the best sky annuli had inner radii between 20--50 pixels and 10--30 pixel widths. The best photometry parameters are compiled in Table~\ref{table:ogle132photparams}.

The final step in the photometric analysis is the identification of potential systematic trends against all available parameters. We examined the out-of-transit baselines of each light curve by fitting linear trends with respect to airmass, seeing, telescope azimuth, (x,y) CCD pixel location, and time. The 20080419 transit light curve showed no significant trends, while the other light curves showed trends against telescope azimuth (20080424), seeing (20090424), and airmass (20080511, 20090207, 2009031, 20090511). With each trend removed, there is a possibility of introducing error into the light curve, though we note that three of our light curves are within 20\% of the photon noise limit.

\begin{figure}
\begin{center}
\includegraphics*[scale=0.45]{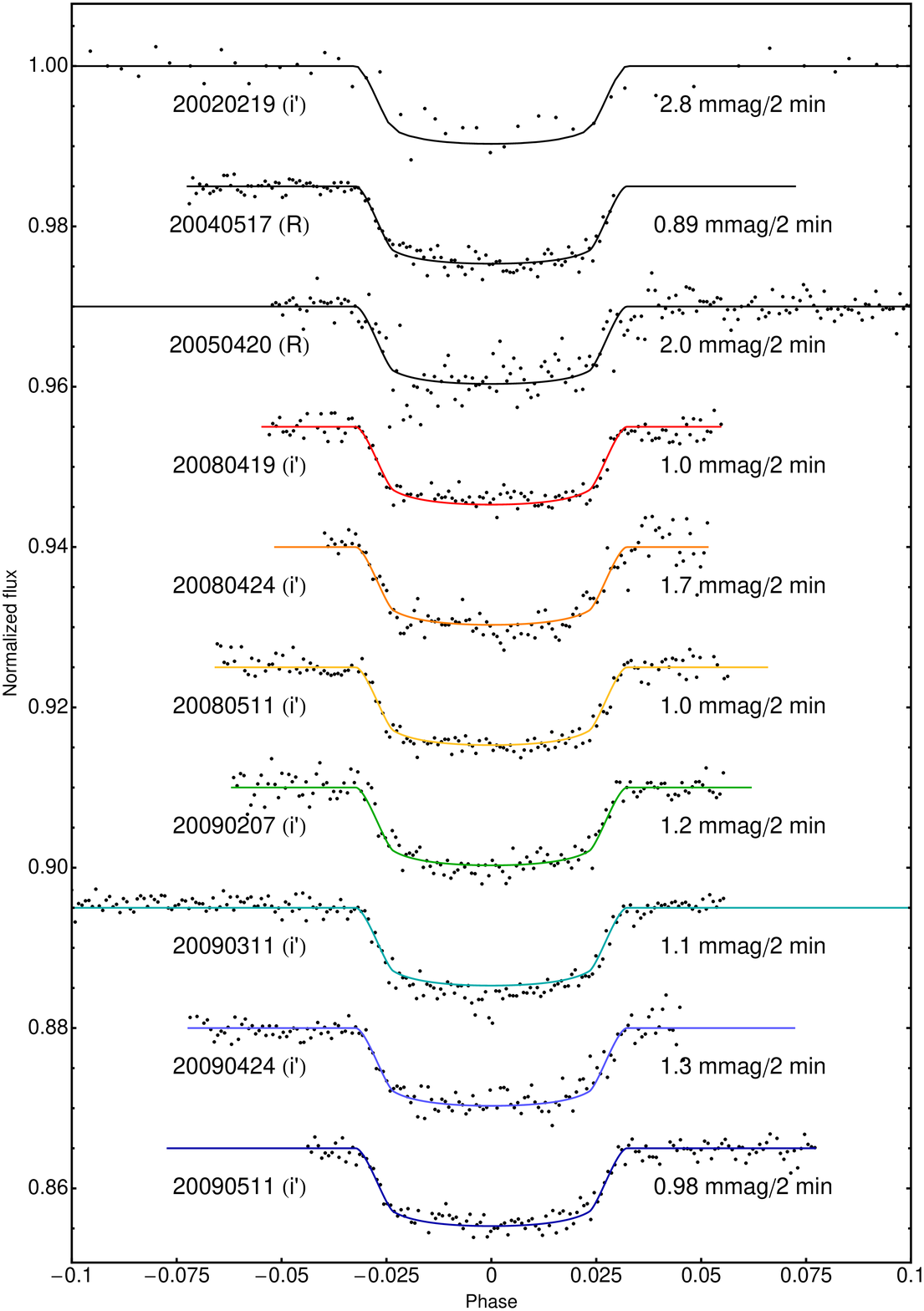}
\end{center}
\caption{Ten transits of OGLE-TR-132b. Normalized flux for all available light curves are plotted vs. orbital phase, with each transit shifted vertically by 0.015 for clarity. Though the full data was used for fits, the data displayed has been binned to 2 minutes (except for the OGLE composite curve 20020219, which was fit with 5 minute bins). The top three transits are taken from \citet{Udalski2003} and \citet{Gillon2007}, while the rest of the transits are new. The joint model fit (solid lines) were calculated using the parameter values in Table~\ref{table:ogle132mcmc}; the stated standard deviations are the residuals of each transit from the joint model fit.}
\label{fig:ogle132transits}
\end{figure}

We now briefly describe the previously published light curves that we used in our analysis. All ten transit light curves are shown in Figure~\ref{fig:ogle132transits}, where we have averaged the data in 120 s bins for representational clarity. The full unbinned light curves are available electronically, and are excerpted in Table~\ref{table:ogle132data}.

\subsection{Literature light curves}

Two $R$-band light curves were reported by \citet{Moutou2004} and \citet{Gillon2007}, referred to throughout as 20040517 and 20050420. These light curves contain 274 and 270 data points, respectively, with samplings of $\sim$60 s and 82 s per exposure (including readout time). The 20050420 dataset includes baseline before and after the eclipse, while the 20040517 light curve ends right before egress.  The average photometric precision of the 20050420 data is 2.0 mmag per 120 s bin, while the 20040517 data, with more stable photometry, has an average dispersion of 0.9 mmag per 120 s time bin. 

Although the original OGLE survey data is much less precise than the other light curves, we include it in our analysis in order to test whether its parameters are consistent with the more recent data. The OGLE discovery curve, denoted 20020219, was observed in the $I$-band and combines a year's worth of full and partial transits \citep{Udalski2003}. To produce the light curve shown in Figure~\ref{fig:ogle132transits}, we phase-folded the data using the orbital period $P = 1.689868$ days \citep{Gillon2007}, and then binned the data over five minutes intervals to decrease the point-to-point scatter before fitting. Due to the faintness of the target and the smaller telescope diameter, we note two caveats for the 20020219 transit data: (a) the photometric error is much larger than any of the other nine light curves, so this transit has little effect on the parameters that result from a noise-weighted fit (described in the next section), and (b) our fitted midtime is 6.5 minutes earlier than the value reported (without errors) in \citet{Udalski2003}, though this discrepancy is also comparable to our revised errors, which are about 6 minutes after accounting for correlated noise.  

\section{Model and light curve fits}
\label{section:fitting}

We fit all ten available light curves jointly, to minimize potential systematics in the individual light curves and to derive average parameters of the OGLE-TR-132b transit with high precision. We also fit each curve (except the low-precision curve 20020219) individually to search for parameter variations over time. 

The light curves were fit in the same manner described in \citet{Adams2010a, Adams2010b}. We used the \citet{Mandel2002} algorithm to generate analytical models, and the basic white-noise model-fitting code described in \citet{Carter2009}, not including the wavelet analysis. We used the same limb darkening quadratic law in Equation 1 of \citet{Adams2010a}, with the initial limb darkening coefficients set to the values computed by \citet{Claret2000, Claret2004} for each filter. The OGLE survey data were observed in $I$-band, but their precision is insufficient to constrain the limb darkening, so we have assumed that the shape of this transit follow the same limb darkening law as the Sloan $i'$ transits. The initial values of the limb-darkening coefficients $u_{1}$ and $u_{2}$ for each filter were obtained using the {\it jktld} routine by \citet{Southworth2008}, assuming the parameters of the host star derived by \citet{Bouchy2004}, $T_{*} = 6411$ K, $\log{g}= 4.86$, $[M/H] = 0.0$ and $V_{micro} = 2$ km s$^{-1}$ (note that although these values differ slightly from those of \citet{Gillon2007} the slight difference in initial limb darkening parameters has negligible impact on the final fitted model parameters). Throughout we fixed the quadratic term of the limb darkening law, $u_{2}$; the linear term, $u_{1}$, was also fixed in the individual light curve fits. The orbital period of the planet was fixed to P = 1.689868 days, after testing confirmed this parameter has little effect on the model, and we assumed throughout that $M_{*} = 1.26 \pm 0.03~M_{\odot}$, $R_{*} = 1.34 \pm 0.08~R_{\odot}$, and $M_{p} = 1.14 \pm 0.12 ~M_{J}$, based on the analysis by \citet{Gillon2007}. The free parameters were the linear limb darkening coefficient, $u_{1}$, the planet-star radius ratio, $k$, the inclination, $i$, the semimajor axis in stellar radii, $a/R_{*}$, the out-of-transit baseline flux, $F_{OOT}$, and the transit midtime, $T_{C}$.

\begin{figure}
\begin{center}
\includegraphics*[scale=0.2]{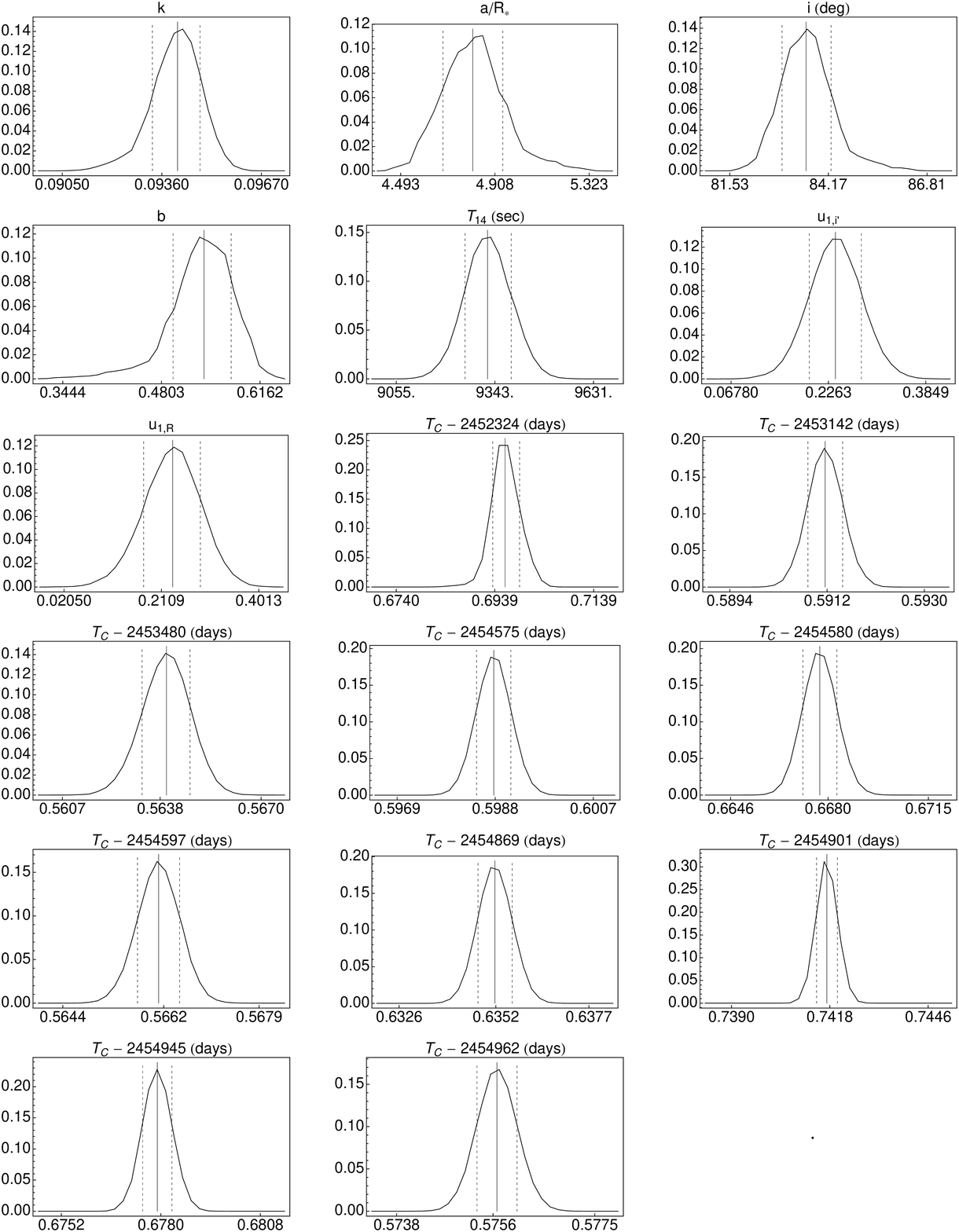}
\end{center}
\caption{Parameter distributions for joint fit to ten transits of OGLE-TR-132b. Smoothed histogram of normalized parameter distributions, from which the parameters in Table~\ref{table:ogle132mcmc} are derived. The solid line is the median value (which is very close to the mean value in all cases). The dashed lines show the 68.3\% credible interval. These values were calculated for 2,850,000 links. 
\label{fig:ogle132dist}}
\end{figure}

To determine the best fit value and error of each parameter, we used a Markov Chain Monte Carlo (MCMC) method as described in \citet{Carter2009}. We first fit each light curve independently and used the resultant reduced $\chi^2$ of each fit as a weighting factor for each light curve. We then fit all ten light curves simultaneously, assuming common values of $k$, $i$, $a/R_{*}$, and  $u_{1,x}$ (where $x$ indicates the filter), and independent values of $F_{OOT}$ and $T_{C}$ for each light curve. We combined three MCMC chains of 950,000 links each, discarding the first 50,000 links of each chain to avoid biasing the solution towards the input values, and produced the distributions shown in Figure~\ref{fig:ogle132dist}. In order to confirm that the chains had truly converged, we calculated the Gelman-Rubin statistic for each parameter, and found that it was never greater than 1.1, indicating acceptable convergence.

In Table~\ref{table:ogle132mcmc}, we report the median value of each parameter distribution, where the error bars correspond to the 68.3$\%$ credible intervals of the distributions. To account for the effect of correlated noise, the errors have been increased according to the residual permutation method as previously discussed in \citet{Adams2010a, Adams2010b}; with this factor, the noise on our transits is 1.2 to 1.9 times greater than the Poisson noise. Table~\ref{table:ogle132mcmc} also reports two derived quantities, the value of the impact parameter, $b$, and the total duration of the transit, $T_{14}$, defined as the time from the start of ingress to the end of egress. 

\begin{figure}
\begin{center}
\includegraphics*[scale=0.3]{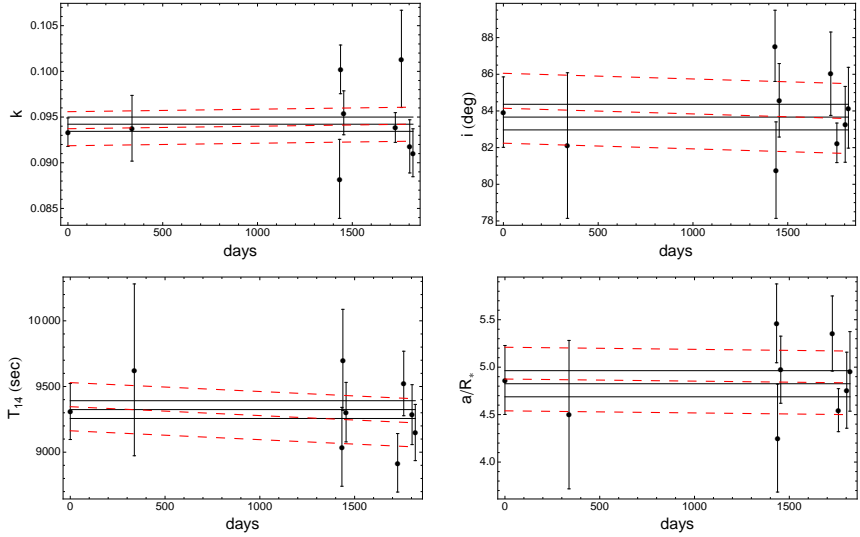}
\end{center}
\caption{Parameter variation of nine individual transits of OGLE-TR-132b, based on individual MCMC fits (Table~\ref{table:ogle132indfits}), for (clockwise, from top-left): $k$, $i$, $a/R_*$, and $T_{14}$. The OGLE composite curve 20020219 has been omitted since it was not fit independently. The values derived from the joint MCMC fit to all transits (Table~\ref{table:ogle132mcmc}) are plotted as solid black lines with $\pm1\sigma$ errors. The dashed red lines indicate the best sloped fit with $\pm1\sigma$ errors. All fits are consistent with no parameter variation during the time period examined.}
\label{fig:ogle132paramvar}
\end{figure}

To explore potential parameters variations over time, we ran additional MCMC fits on the individual transit light curves. We excluded the OGLE survey composite transit, 20020219, because the photometric precision is much worse than the rest of the data, and the model for this light curve is not well constrained without fixing several parameters. For the nine light curves we examined individually, the results for four parameters, $k$, $i$, $T_{14}$ and $a/R_{*}$, are summarized in Figure~\ref{fig:ogle132paramvar}. We find no evidence of variation of any of those parameters, within the error bars, during the five year time span covered by those observations.

The planetary radius of a transiting planet can be difficult to determine in a crowded field, since even very faint comparison stars may dilute the depth of the measured transit if not carefully accounted for. The two literature light curves published in \citet{Gillon2007} were both created using image subtraction, and that work was careful to note that the choice of background normalization factor can greatly affect the depth obtained. All seven of the new transits were observed during excellent seeing, which allowed us to use aperture photometry despite the crowded field. (We note that a star that is about 3.5 magnitudes fainter is located 1.3", or 36 unbinned pixels, from our target; this is far enough outside of our chosen apertures that its effect would be much less than the millimag precision of our light curves.) 

Despite this caveat, we note that the radius we obtain, $R_p = 1.23 \pm 0.07$ $R_J$, is very similar to that of \citet{Gillon2007}, $R_p = 1.18 \pm 0.07$ $R_J$. It also agrees well with the radius derived by \citet{Southworth2008}, $R_p = 1.25 \pm 0.16$ $R_J$, which was derived from a reanalysis of the higher-precision light curve, 20040517, from \citet{Gillon2007}. Note that the error in the \citet{Southworth2008} reanalysis, which accounts for correlated noise, is larger than the \citet{Gillon2007} error; this is because the transit is missing the end of egress, which makes it particularly vulnerable to correlated noise. With the addition of seven full light curves, the planetary radius error can be determined much more precisely, and is currently dominated by the uncertainties in the radius of the star, $R_{*} = 1.34 \pm 0.08$ $R_{\odot}$; without the stellar error, the radius error would have been 0.01 $R_{J}$, seven times smaller.

\section{Timing}
\label{section:timing}

We now turn to an analysis of the timing of each light curve. Each transit midtime, $T_C$, derived in the previous section is listed in Table~\ref{table:ogle132mcmc}. Times for the seven new transits have been converted to Barycentric Julian Days using the Barycentric Dynamical Time standard ($BJD_{TDB}$), as recommended by \citet{Eastman2010}. In addition, we have converted the three literature datasets to the same $BJD_{TDB}$ time standard by adding the UTC-TT correction, after confirming with the authors that none of those data had previously accounted for it (M. Gillon and A. Udalski, private communication).

During the course of this analysis, we noted that the original $T_C$ published for the OGLE survey dataset (20020219) does not agree with the more recent transits, prompting us to refit the light curve.  Our joint fit of all ten transits, which provides high-precision constraints on the light curve shape of the OGLE composite curve, gives a midtime of $T_C = 2452324.69684 \pm 0.00284$ BJD$_{TDB}$, 331 seconds earlier than the published time \citep[$2452324.70067$ BJD$_{UTC}$, with errors assumed to be about 0.001,][]{Udalski2003}. After correcting for the UTC-TT offset (64.184 s) that timing difference increases to 395 seconds. Our newly calculated transit midtime agrees well with the midtimes of the other nine transits, though with quite large error bars: the formal error is 245 s, but after accounting for the effects of strong correlated noise, we adopt an error of 343 s, very similar in magnitude to the timing shift, although the ultimate cause remains unclear. We use our recalculated midtime value for 20020219 in the remainder of our analysis, while noting that its large error bars mean it has little effect on the weighted fits. 

\begin{figure}
\begin{center}
\includegraphics*[scale=0.5]{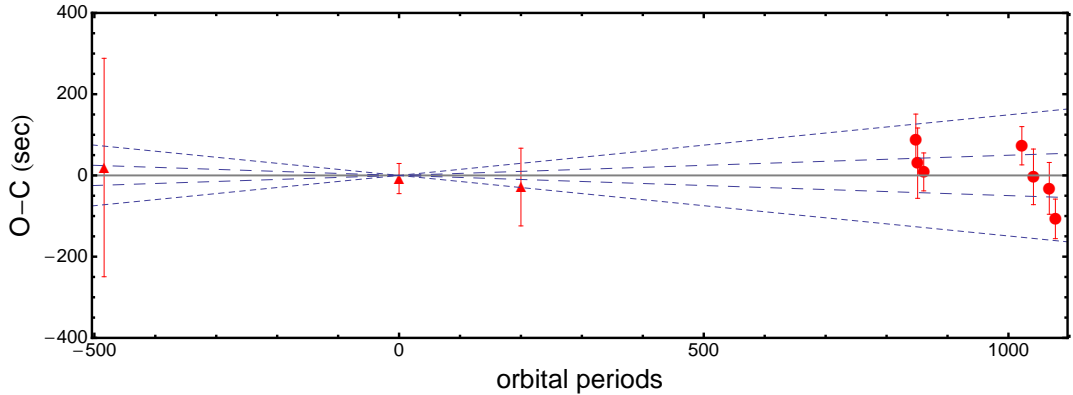}
\end{center}
\caption{Observed minus calculated midtimes for OGLE-TR-132b. Timing residuals using the new ephemeris (Equation~\ref{ogle132eqn1}), calculated using all ten transit epochs. The solid line represents zero deviation from expected time of transit, while the dashed lines represent the $1\sigma$ and $3\sigma$ errors on the calculated orbital period, indicating the slopes that result for a mis-determined period. }
\label{fig:ogle132ominusc}
\end{figure}

Assuming a constant period, we derived a new transit ephemeris equation from a weighted fit to all ten transits, with the form:
\begin{equation}
T_{C}(N) = 2453142.59199(53) [BJD_{TDB}] + 1.68986531(64) N,
\label{ogle132eqn1}
\end{equation}
where $T_C$ is the predicted midtime of transit and $N$ is the number of orbital periods since the reference midtime. The values in parentheses are the errors in the last two digits. This fit, represented in Figure~\ref{fig:ogle132ominusc} by the solid line, has a reduced $\chi^2 = 1.2$, indicating a good fit; the errors reported in the equation have been rescaled slightly, by a factor of $\sqrt{1.2} = 1.1$, to account for the difference from $\chi^2 = 1.0$.  Figure~\ref{fig:ogle132ominusc} shows the timing residuals, or observed minus calculated time ($O-C$), for each transit with respect to the new ephemeris equation above.

The new orbital period of OGLE-TR-132b is five times more precise than the most recent value, reported by \citet{Gillon2007}. Such improvement is due to the longer transit follow-up time baseline available for this system. Nonetheless, the new orbital period differs by less than 0.2 seconds from the one derived by \citet{Gillon2007}, well within the 1 $\sigma$ errors. Additionally, only one individual transit deviates by more than 2 $\sigma$ from the fit (20090511, $-108 \pm 49$ s, or 2.2$\sigma$), with most of the rest within 1 $\sigma$ of the best fit line. There is thus no reason to conclude that there are any significant timing variations in this system. 

Using the lack of significant timing variations, we can attempt to constrain the rate of orbital decay, and provide upper limits on the masses of potential companion planets.

\subsection{Limits on orbital decay}

Recent calculations by \citet{Levrard2009} predict a lifetime for OGLE-TR-132b of only 52 Myr, assuming the tidal dissipation factor for the star of $Q_{*} = 10^6$. (For $Q_{*} = 10^5-10^{10}$, the lifetime would be 5.2 Myr to 520 Gyr.) Recent observations of a different hot Jupiter, OGLE-TR-113b, hint that its orbital period may be decreasing by $\dot{P}_{113b}=-60\pm15$ ms yr$^{-1}$ \citep{Adams2010b}, which, if confirmed, leads to an estimate of $Q_{*} = 1.6\times10^4$ using the \citet{Levrard2009} calculations of tidal dissipation. If we assume the same value of the $Q_*$ for OGLE-TR-132, the remaining lifetime of the planet scales to only 0.8 Myr. 

To estimate the predicted rate of decay per year, we define the lifetime of the planet as the time that takes for it to reach the Roche lobe of the star. We estimated the radius of the Roche lobe using the analytical approximation derived by \citep{Paczynski1971}, $r_{RL} = 0.462 ~ a~ (M_p / (M_p+M_{*}))^{1/3}  = 0.014$ AU for OGLE-TR-132b, assuming a circular orbit.  The corresponding total decrease in period is 1.16 days, so assuming a constant rate of decrease, the orbital decay rate is 2 ms yr$^{-1}$ if $Q_{*} = 10^6$, or a much faster 140 ms yr$^{-1}$ if $Q_{*} = 1.6\times10^4$. This is, perhaps coincidentally, close to the best fit value of a decreasing-period model if the period of OGLE-TR-132b is decaying ($\dot{P}_{132b}=-125\pm80$ ms yr$^{-1}$, with reduced-$\chi^2=0.8$), but with a significance of only 1.5~$\sigma$ this value is entirely consistent with zero period change. Since the constant-period ephemeris in Equation~\ref{ogle132eqn1} fits the data so well, there is no reason to conclude that this system is undergoing orbital decay, within the limits of our precision and length of observations.

Additional observations of OGLE-TR-132b over the next decade will be important to refine the limits on orbital decay.

\subsection{Limits on perturber mass}

The lack of significant TTVs for OGLE-TR-132b over the seven year time span of the compiled observations indicate the absence of other massive, nearby companions in this system. However, smaller planets can be still hiding within the 50-100 s dispersion of our TTVs. In this section we report upper limits to the mass and orbital separation of hypothetical perturbing planets, which would cause TTV signals with amplitudes smaller than our detection limit. Those upper limits were computed using an implementation of the \citet{Steffen2005} algorithm, as implemented by D. Fabrycky (2009, private communication). 

\begin{figure}
\begin{center}
\includegraphics*[scale=0.5]{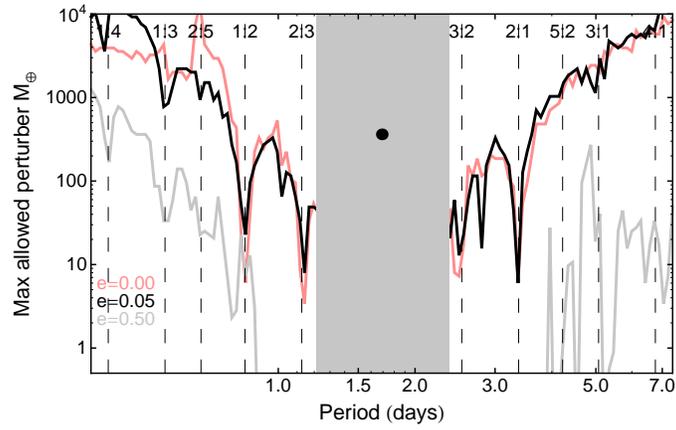}
\end{center}
\caption{Upper mass limit on additional planets with initial eccentricities $e_c=0.0$ (red), $e_c=0.05$ (black) and $e_c=0.5$ (gray). The central point shows OGLE-TR-113b. Near the 1:2 and 2:3 mean-motion resonances, objects as small as 5-10~$M_{\oplus}$ would have been detectable. The shaded grey region shows the instability region for a $1~M_{\oplus}$ companion in a circular orbit, following \citet{Barnes2006}.}
\label{fig:ogle132mass}
\end{figure}

Figure~\ref{fig:ogle132mass} shows the maximum allowed mass for a potential perturbing planet over a range of orbital periods. These calculations were performed using only the seven new transits and the best literature transit (20040517), due to the higher errors on the other two literature points. For simplicity, we assumed that the perturber is coplanar with OGLE-TR-132b, and explore several orbital eccentricity values. The most conservative constraints are place by objects with the lowest eccentricities, $e=0$ and $e=0.05$. Objects with higher eccentricities (represented here by the $e=0.5$ case), might be expected if the object were scattered by Kozai migration \cite[e.g.,][]{Fabrycky2007,Triaud2010}, but near OGLE-TR-132b such objects tend to be much more unstable, and to produce larger hypothetical TTV signals (and correspondingly lower mass limits). A gray shaded region in Figure~\ref{fig:ogle132mass} shows the approximate unstable region for an Earth-mass object on a circular orbit, following \citet{Barnes2006}. We examined a range of orbital period ratios from 1/4 and 4 times that of OGLE-TR-132b (or 0.4 to 7.4 days), divided into 51 bins on a logarithmic scale. For each period, we increased the mass of the perturber from an initial value of 0.0001 $M_{J}$, increasing the mass by a factor of 1.1 with each step until we find the mass that would produce a TTV equivalent to the 3$\sigma$ confidence level of our results, which we report as the upper limit on the perturber's mass.

The mass constraints placed are strongest at orbital distances near low-order mean motion resonances with OGLE-TR-132b. In particular, we can discard the presence of planets as small as 5-10 $M_{\oplus}$ in the external 2:1 and 3:2 and internal 1:2 and 2:3 resonances in circular or nearly circular orbits. More eccentric objects have even smaller limits mass limits outside of the 1:2 and 2:1 resonances, and are unstable on closer orbits. No significant constraints can be placed on near-circular objects at orbital distances outside the 2:1 resonance, at which period perturbers of any mass (including stellar) are essentially unconstrained by the timing. Such objects are better constrained by other methods, such as radial velocity, direct detection, and dynamical stability arguments.

Finally, neither the precision of our light curves in Figure~\ref{fig:ogle132transits} nor the precision of our timing residuals in Figure~\ref{fig:ogle132ominusc} are sufficient to constrain the presence of moons around OGLE-TR-132b. The best achieved precision on any light curve was about 1 mmag on the photometric depth, 37 s on the transit midtime, and 213 s on the transit duration, whereas an Earth-sized planet located within the Hill radius of the planet would produce transit signals of 0.05 mmag, a TTV smaller than a few seconds, and a transit duration variation, or TDV, of a few seconds \citep{Kipping2009a, Kipping2009b}.

\section{Summary and Conclusions}
\label{section:conclusions}

We have measured seven new transits of the hot Jupiter OGLE-TR-132b, which triples the number of transits available for this planet to date. The new transits have timing precisions between 47 and 87 seconds and photometric precision from 1-2 mmag in two minute bins. The updated observational dataset of OGLE-TR-132b spans seven years (2002--2009) and a total of ten transits after combining the new transits presented in this paper with three previously published transit epochs.

We find no evidence of transit timing variations with amplitudes larger than $-108 \pm 49$ s, which allows us to place significant constraints to the presence of additional close-in planets in the system. In particular, near both the internal and external 2:1 and 3:2 mean motion resonances, objects as small as 5-10 Earth masses would have been detectable if they existed.

We also tested for orbital period decay, with the best fit rate of period decrease, $\dot{P}_{132b}=-125\pm80$ ms yr$^{-1}$, consistent with no change. The predicted orbital period decay rate for this planet is 20 ms yr$^{-1}$ if $Q_{*} = 10^5$, and 140 ms yr$^{-1}$ if $Q_{*} = 1.6 \times 10^4$ as suggested for the star OGLE-TR-113 \citep{Adams2010b}, so additional transits with better timing precision ($<30$ s) or a significantly longer time baseline are needed to search for slower rates of period change.

Finally, we have measured a revised radius value for the planet, $R_p = 1.23 \pm 0.07$ $R_J$, which is fully consistent with the radius derived by \citet{Southworth2008} from a reanalysis of the \citet{Gillon2007} and \citet{Moutou2004} data. We used simple aperture photometry to extract all transit light curves, so the depth of our transits should not be affected by the potential normalization problems found when using image differencing techniques, as reported for this same planet by \citet{Gillon2007}. Our planet radius provides an independent confirmation, within errors, of the transit depth obtained by \citet{Gillon2007} using image deconvolution techniques.

\acknowledgements

E.R.A. received partial support from NASA Origins grant NNX07AN63G. M.L.M. acknowledges support from NASA through Hubble Fellowship grant HF-01210.01-A/HF-51233.01 awarded by the STScI, which is operated by the AURA, Inc. for NASA, under contract NAS5-26555. We thank Paul Schechter, Brian Taylor, and the Magellan staff for their invaluable assistance with the observations, Josh Carter and Dan Fabrycky for the use of their software, and Dave Charbonneau and Josh Winn for helpful discussions.

\clearpage




\newpage

\clearpage


\begin{deluxetable}{llllllllll}
\tablewidth{0pt}
\tabletypesize{\scriptsize}
\tablecaption{Observational parameters for seven transits of OGLE-TR-132b}
\tablehead{
Transit		& Frames\tablenotemark{a}	& Exp. Time	& Binning	& Readout 			& Airmass 	& Time range observed		& Atm. stability	& Seeing	\\
(UT)			& 							& (sec)		& 		& (sec)				&	 		&	 	&	 	}
\startdata
20080419		& 227 (0)						& 60			& 1x1	& 5					& 1.2--1.4	 	& 4.5 hrs around midtransit	&  Stable		& $0.\arcsec4$--$0.\arcsec5$	\\
20080424		& 160 (0)						& 60, 120		& 1x1	& 5					& 1.2--1.8 		& 3 hrs (25 min before ingress)	&Variable		& $0.\arcsec 6$--$0.\arcsec 7$ \\
20080511		& 136 (5)	\tablenotemark{b}		& 120		& 1x1	& 5					& 1.2--1.5	 	& 5 hrs around midtransit	 	& Stable		& $0.\arcsec 5$--$0.\arcsec 7$\\
20090207		& 1190 (10)\tablenotemark{c}		& 10, 15		& 2x2	& \tablenotemark{d}	& 1.2--2.1	 	& 5 hrs around midtransit	 	& Variable& $0.\arcsec4$--$0.\arcsec7$ \\
20090311		& 2322 (0)					& 10, 15		& 2x2	& \tablenotemark{d}	& 1.2--1.6	 	& 8 hrs (4 hrs before ingress)	&Very stable	&  $0.\arcsec6$ \\
20090424		& 476 (0)						& 20-75		& 1x1	& \tablenotemark{d}	& 1.2--1.6	 	& 5 hrs (30 min before ingress)	&Variable		& $0.\arcsec3$--$0.\arcsec4$ \\
20090511		& 440 (0)						& 35, 40		& 2x2	& \tablenotemark{d}	& 1.2--1.8	 	& 5 hrs (35 min before ingress)	& Stable		& $0.\arcsec5$ \\
\enddata
\tablenotetext{a} {Number of frames used (additional frames that were discarded).}
\tablenotetext{b} {Discarded due to tracking problem.}
\tablenotetext{c} {Discarded due to low counts.}
\tablenotetext{d} {Readout is a few miliseconds in frame transfer mode.}
\label{table:ogle132obsparams}
\end{deluxetable}

\begin{deluxetable}{llllllllll}
\tablewidth{0pt}
\tabletypesize{\scriptsize}
\tablecaption{Photometry parameters for OGLE-TR-132b}
\tablehead{
Transit		& Comp. Stars	& Aperture\tablenotemark{a}	& Sky radius, width	& Slope removed 	& Scatter\tablenotemark{b}	& Noise factor\tablenotemark{c}	\\
(UT)			&	 		&  (pixels)					& (pixels)		&				& (mmag)					&} 
\startdata
20080419		& 7	& 13			& 30, 20	& None						& 1.5		& 1.3		 \\
20080424		& 4	& 19			& 40, 30	& $+0.013\%$ \tablenotemark{d}	& 2.2		& 1.9		\\
20080511		& 3	& 26			& 50, 20	& $-1.2\%$  \tablenotemark{e}		& 1.3		& 1.4		\\
20090207		& 2	& 15			& 30, 10	& $+0.1\%$ \tablenotemark{e}		& 3.9		& 1.6		\\
20090311		& 2	& 12			& 30, 10	& $-0.3\%$ \tablenotemark{e}		& 2.9		& 1.2		\\
20090424		& 3	& 15			& 30, 10	& $-0.052\%$  \tablenotemark{f}	& 2.0		& 1.2		\\
20090511		& 8	& 9			& 20, 10	& $+0.12\%$ \tablenotemark{e}	& 1.8		& 1.2		\\
\enddata
\tablenotetext{a} {Radius around star}
\tablenotetext{b} {Frame-by-frame scatter on out-of-transit flux (not corrected for exposure time).}
\tablenotetext{c} {Factor by which scatter exceeds Poisson noise estimate.}
\tablenotetext{d} {Trend removed against telescope azimuth, in units of flux deg$^{-1}$	.}
\tablenotetext{e} {Trend removed against object airmass, in units of  flux X$^{-1}$.}
\tablenotetext{f} {Trend removed against seeing in pixels, in units of flux pixels$^{-1}$.}
\label{table:ogle132photparams}
\end{deluxetable}

\begin{deluxetable}{l l l l }
\tablewidth{0pt}
\tabletypesize{\scriptsize}
\tablecaption{Flux values for seven transits of OGLE-TR-132b}
\tablehead{ Mid-exposure (UTC)\tablenotemark{a}	& Mid-exposure (BJD)	& \textrm{Normalized stellar flux}  		& RMS error}
\startdata
2454575.506679	&2454575.510331	&0.9996423	&0.00122 \\
2454575.507433	&2454575.511084	&1.001028	&0.00122 \\
2454575.508183	&2454575.511834	&0.9992174	&0.00122\\
2454575.508936	&2454575.512588	&1.000828	&0.00122\\
2454575.509689	&2454575.513341	&1.000836	&0.00122\\
\nodata
\enddata     
\tablenotetext{a} {This table is available in its entirety in a machine-readable form in the
online journal. A portion is shown here for guidance regarding its form and content.}
\label{table:ogle132data}
\end{deluxetable}


\begin{deluxetable}{l c c  c }
\tablewidth{0pt}
\tabletypesize{\scriptsize}
\tablecaption{Transit parameters for OGLE-TR-132b (jointly fit)}
\tablehead{Parameter		& Median value\tablenotemark{a}}
\startdata
\emph{Fitted Parameters} \\
$k$						& $0.0941 \pm 0.0014$  \\
 $a/R_*$			 		& ~~$4.81 \pm 0.16$~~ 		 \\
 $i$ (deg)			 		& ~~$83.6 \pm 0.7$~~~		\\
 $u_{1,i'}$					&~~ $0.24 \pm 0.04$~~			 \\
 $u_{2,i'}$ 				& 0.19 (fixed)	 \\
  $u_{1,R}$				& ~~$0.23 \pm  0.06$~~	\\
 $u_{2,R}$				& 0.38 (fixed)	 \\
$T_C - 2452324$ 			&~~ ~~~~~~$0.69684 \pm 0.00400$ (343 s)	\\
$T_C - 2453142$ 			& ~~~~~~~~$0.59191 \pm 0.00043$ ~(37 s)	\\
$T_C - 2453480$ 			& ~~~~~~~~$0.56473 \pm 0.00110$ ~(94 s)	\\
$T_C - 2454575$ 			& ~~~~~~~~$0.59877 \pm 0.00074$ ~(64 s)	\\
$T_C - 2454580$			& ~~~~~~~~$0.66772 \pm 0.00100$  ~(87 s)	 \\
$T_C - 2454597$			& ~~~~~~~~$0.56611 \pm 0.00054$  ~(47 s)	\\
$T_C - 2454869$ 			& ~~~~~~~~$0.63518 \pm 0.00055$ ~(47 s)	\\
$T_C - 2454901$			& ~~~~~~~~$0.74172 \pm 0.00079$ ~(68 s)	\\
$T_C - 2454945$ 			& ~~~~~~~~$0.67790 \pm 0.00075$ ~(64 s)	\\
$T_C - 2454962$ 			& ~~~~~~~~$0.57567 \pm 0.00057$  ~(49 s)	\\
\\  
\emph{Derived Parameters}\\
 $b$				 			& ~~$0.53\pm 0.05$~~   \\
 $T_{14}$ (sec)		 			& ~$9322\pm111$~ ~  \\
 $R_p$ ($R_J$)\tablenotemark{b}  	& ~$1.227\pm 0.074$~  \\
 $a$ (AU)\tablenotemark{b} 	 	& ~$0.030\pm 0.002$~   \\
\enddata     
\tablenotetext{a} {Median value of parameter distribution with errors increased from the $68.3\%$ credible interval by residual permutation method (see Section~\ref{section:fitting}).}
\tablenotetext{b} {Assuming $R_*=1.34 \pm 0.08 ~R_{\odot}$ \citep{Gillon2007} and using $R_J=71,492$ km.}
\label{table:ogle132mcmc}
\end{deluxetable}

\begin{deluxetable}{l c l c l c l c l}
\tablewidth{0pt}
\tabletypesize{\scriptsize}
\tablecaption{Individual transit parameters for OGLE-TR-132b (independently fit)}
\tablehead{
Transit &$k$\tablenotemark{a} & $f$\tablenotemark{b} &$T_{14}$\tablenotemark{a} & $f$\tablenotemark{b} &$a/R_*$\tablenotemark{a} & $f$\tablenotemark{b} &$i$\tablenotemark{a} & $f$\tablenotemark{b} }
\startdata		
20040517 & $0.0933 \pm 0.0016$ & \nodata		& $9310 \pm 213$ & 1.1 		& $4.86 \pm 0.36$ & \nodata	 & $83.9 \pm 1.9$ & \nodata \\
20050420 & $ 0.0938 \pm 0.0036$ & \nodata	 	& $9627 \pm 654$ & \nodata 	& $4.50 \pm 0.78$ & \nodata	 & $82.1 \pm 4.0$ & \nodata \\
20080419& $ 0.0882 \pm 0.0043$ & 4.1	 		& $9041 \pm 300$ & 2.9 		& $5.46 \pm 0.42$ & 1.9 		& $87.6 \pm 1.9$ & 1.2 \\
20080424 & $0.1002 \pm 0.0027$ & \nodata	 	& $9703 \pm 384$ & 1.1 		& $4.25 \pm 0.57$ & 1.4 		& $80.8 \pm 2.6$ & 1.4 \\
20080511 & $0.0955 \pm 0.0024$ & 1.6			 & $9305 \pm 226$ & 1.3 		& $4.97 \pm 0.35$ & \nodata 	& $84.6 \pm 2.0$ & \nodata \\
20090207 & $0.0939 \pm 0.0016$ & \nodata 		& $8919 \pm 223$ & 1.3 		& $5.35 \pm 0.40$ & \nodata 	& $86.0 \pm 2.3$ & \nodata \\
20090311 & $0.1014 \pm 0.0053$ & 4.6 			& $9523 \pm 245$ & 1.6 		& $4.55 \pm 0.23$ & \nodata 	& $82.3 \pm 1.1$ & \nodata \\
20090424 & $0.0918 \pm 0.0029$ & 1.7		 	& $9286 \pm 228$ & 1.1 		& $4.76 \pm 0.40$ & \nodata 	& $83.3 \pm 2.1$ & \nodata \\
20090511 & $0.0911 \pm 0.0026$ & 1.6 			& $9150 \pm 214$ & 1.0 		& $4.96 \pm 0.42$ & \nodata 	& $84.2 \pm 2.2$ & \nodata \\
\enddata
\tablenotetext{a} {Formal individual MCMC fit value and error (scaled upward by factor $f$ in adjacent column).}
\tablenotetext{b} {Factor by which the error in the previous column has been increased based on the residual permutation method; no value is given if the error from residual permutation is smaller than the MCMC fit error.}
\label{table:ogle132indfits}
\end{deluxetable}

\begin{deluxetable}{l r r c}
\tablewidth{0pt}
\tablecaption{Timing residuals for OGLE-TR-132b}
\tablehead{Transit & Number & $O-C$ (s)&  $\sigma_{O-C}/(O-C)$}
\startdata
20020219		& -484	& $-29 \pm 343$ & -0.1 \\
20040517		& 0 		& $-7 \pm 37$ & -0.2 \\
20050420		& 200	 & $-28 \pm 94$ & -0.3 \\
20080419		& 848	&$ 87 \pm 64$ & 1.4 \\
20080424		& 851 	& $31 \pm 87$ & 0.4 \\
20080511		& 861 	& $8 \pm 47$ & 0.2 \\
20090207 	&1022 	&$ 73 \pm 47$ & 1.6 \\
20090311		&1041 	& $-5 \pm 68$ & -0.1 \\
20090424		&1067 	& $-32 \pm 64$ & -0.5 \\
20090511		&1077 	& $-108 \pm 49$ & -2.2
\enddata
\label{table:ogle132ominuscvalues}
\end{deluxetable}

\clearpage

\end{document}